# Shadows over the speed of light

**J. J. Mareš[1], P. Hubík, V. Špička, J. Stávek, J. Šesták and J. Krištofik**

*Institute of Physics ASCR, v. v. i., Cukrovarnická 10,*
*162 00 Praha 6, Czech Republic*

**Abstract:** In this contribution we are discussing some consequences of the CGPM (1983) definition of meter and especially of giving the speed of light an exact value. It is shown that this act touches the fundamental paradigms, such as the Second postulate of the Special Theory of Relativity (STR), the $c$-equivalence principle and the method of time synchronization. In order to fill the arising logical gaps we suggest among others to weaken the Second postulate of STR to form directly confirmed by experiments and make new measurements of Maxwell's constant with accuracy comparable with that of the speed of light.

## 1. Introduction

Shortly after the discovery of advanced experimental techniques, such as cavity resonance method and laser interferometry in the 1970's, an enormous progress in increasing the accuracy of the measurement of the speed of light was made. Using a 1960-definition of meter in terms of a particular spectral line of krypton-86, and newly constructed laser interferometer, a group at NBS, Boulder, Colorado (1972, [1]), obtained for the speed of light a value $c = 299\,792\,456.2 \pm 1.1$ m/s, which was ~100 times less uncertain than the values accepted previously. As the similar systematic experiments made at that time in competing laboratories provided results of comparable or of even better accuracy, the 15[th] *Conférence Générale des Poids et Mesures* (CGPM) held in 1975 recommended to use for the speed of light a value $c = 299\,792\,458$ m/s. The results of terrestrial measurements of the speed of light performed by various techniques during the period 1907–1974 are summarized in Fig.1. The convergence of measured values of $c$ to a certain constant in two last decades is really remarkable. Due to this very fact and because of inadequacy of the system of units for some metrological experiments the 17th. CGPM (1983) decided also to redefine the meter [2] in the following way: *"The meter is the length of the path travelled by light in vacuum during a time interval* 1/299792458 *of a second."* Consequence of this definition is that the speed of light became an exact constant, namely

$$c = 299\ 792\ 458 \text{ m/s}. \tag{1}$$

---

[1] maresjj@fzu.cz

Apparently, such a far reaching decision was strongly influenced by the general acceptance of the Special Theory of Relativity (STR) according to which the speed of light is a fundamental universal constant preserving its numerical value in all inertial systems. It is an important property of this definition that it is applicable, due to the form of the Lorentz transformations of time and length, not only to the measurements of length in a rest system but in other inertial systems as well. The authors of the reform have thus believed that the improvement of experimental techniques will not affect the value of $c$, but instead, it will allow us a more precise realization of the meter.

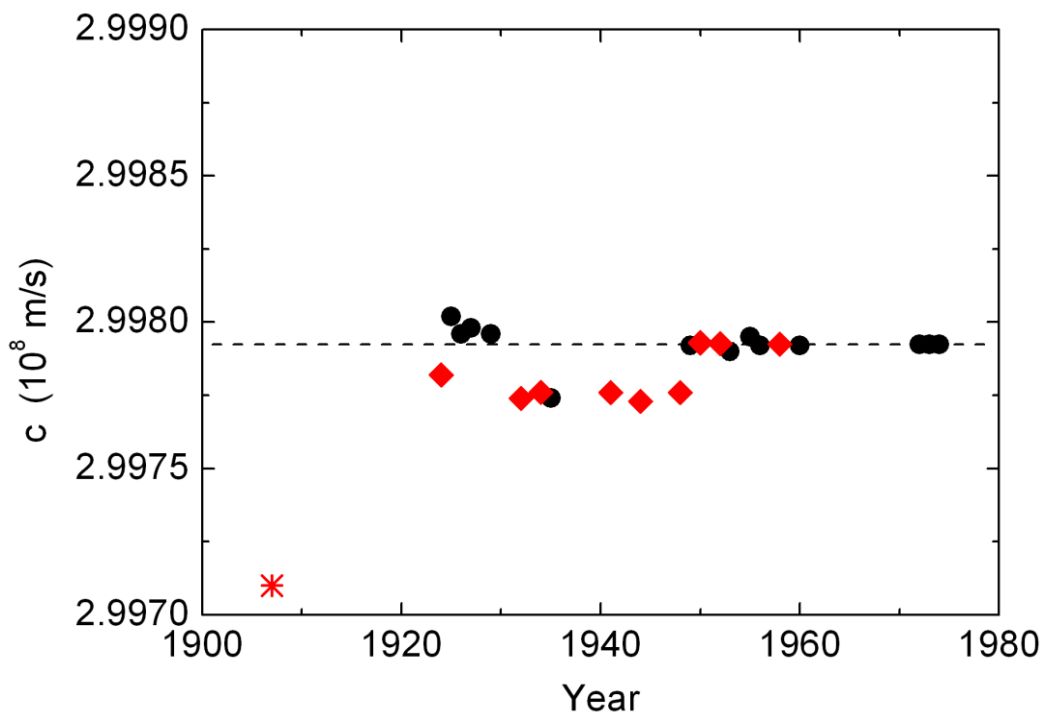

Fig. 1: Results of measurements of speed of light as performed during the period 1907 – 1974. • Optical experiments using visible light, ♦ experiments using electromagnetic waves of lower

frequency. Point ∗ at 1907 corresponds to the measurement of Maxwell's constant $b$ by Rosa and Dorsey [3]. The CGPM (1975) value of $c$ = 299 792 458 m/s is depicted by a dashed line.

Beside quite evident advantages of the introduction of speed of light in the physics and metrology in the form of a fundamental constant having an exact value, we see also some weak points related to such an act. Just the discussion of these controversial weak points, summarized in the following list, is the subject of this paper:

- The kinematic origin of magnetic force and the nature of Maxwell's constant $b$
- Maxwell's equations, $c$-equivalence principle
- Transformation properties of Maxwell's equations, Bessel-Hagen invariants
- Kinematics of light rays and the criticism of the Second postulate of the STR
- CGPM (1983) definition of meter vs. G. de Bray scenario.

## 2. Origin of magnetic forces and Maxwell's constant $b$

The electromagnetic field "*an sich*" is a typical "as if" entity. The very existence of the electromagnetic field apart from the "ponderable" matter is questionable since its experimental investigation without the interaction with material objects is impossible. The electric and magnetic bodies thus behave "as if" there would be something mediating their interactions and to be honest, nobody knows what the electromagnetic field is or what it should be. Putting aside, however, such epistemological questions, let us recall the experimental facts and the theoretical assumptions which are the sources of our knowledge of the electromagnetic field.

The present theory of electromagnetic field, essentially due to J. C. Maxwell, was developed by a generalization of three experimentally established laws, the Coulomb law, Ampère's law (or its equivalent the Biot-Savart law) and the Faraday induction law. Only afterwards experimentally confirmed hypothesis of "current of displacement" was added to these laws. Let us for a while turn our attention to the first couple of these laws. The Coulomb law defines the electrostatic force $F_E$ exerted in an empty space by a point charge $Q_1$ on another one $Q_2$ placed at a distance $R$. Both the point charges are assumed to be in the rest with respect to the observer. Analytical form of the law then reads:

$$F_E = - Q_1 Q_2 / (4\pi\varepsilon_0 R^2). \qquad (2)$$

The proportionality factor $(1/4\pi\varepsilon_0)$ is arbitrary in the sense that it depends exclusively on the choice of system of units. (In our case where the SI system is used, the force is measured in newtons, the distance in meters and the charge in coulombs.) Another important feature of the Coulomb law (2), not apparent at first glance, is that it describes an instantaneous action at a distance (*actio in distans*) along the actual line connecting point charges [4]. Just the same property, i.e. the instantaneous action at a distance along the connecting lines, is given also to a magnetic force between two straight parallel thin wires ("filaments") bearing currents $I_1$ and $I_2$, respectively. Their mutual attraction per length $L$ is controlled by the so called Biot-Savart law, namely

$$F_M = \mu_0 \, I_1 \, I_2 \, L / (2\pi \, R), \qquad (3)$$

where $R$ is the distance between the wires. In a consistent system of units where the current units are defined in terms of charge units, the constants in formulae (2) and (3) cannot be independent. This fact together with another fundamental link between laws (2) and (3) can be elucidated by means of a simple thought experiment. Let us first rewrite the Coulomb law (2) for a system having identical geometry as that used for the formulation of the Biot-Savart law (3). The repulsion between segments of length $L$ of two straight parallel filaments uniformly charged with densities $\gamma_1 = Q_1/L$ and $\gamma_2 = Q_2/L$ which are placed at distance $R$ apart, is given by the formula [5]:

$$F_E = -\, \gamma_1 \, \gamma_2 \, L / (2\pi\varepsilon_0 \, R). \qquad (4)$$

Let us now imagine that the charged system starts to move with the velocity $v$ in the direction parallel to the wires. Obviously, for an observer in the rest-system the charged filaments moving relatively with velocity $v$ represent the currents $I_1 = v \, \gamma_1$ and $I_2 = v \, \gamma_2$. Comparing thus formulae (3) and (4), we obtain for the dimensionless ratio of magnetic and electric force the relation

$$F_M / F_E = -\, \varepsilon_0 \, \mu_0 \, v^2. \qquad (5)$$

Immediate physical interpretation of our thought experiment and formula (5) is the following: The magnetic force is nothing but the electric force observed from a relatively moving system. Since the ratio $F_M / F_E$ is dimensionless, it is possible to introduce a normalizing factor having physical dimension of the speed, namely

$$b = 1/\sqrt{(\varepsilon_0\mu_0)}. \qquad (6)$$

This quantity, sometimes called Maxwell's constant, enables one to express the total force of electromagnetic origin $F_{EM} = F_E + F_M$ in a compact form

$$F_{EM} = F_E (1 - v^2/b^2), \qquad (7)$$

from which it follows that the magnetism is with respect to electrostatic interaction only a second order effect. In accordance with formula (6), one of the quantities $\varepsilon_0$ and $\mu_0$ can be chosen arbitrarily, while the other one or $b$ should be determined experimentally. For example, putting in the SI system of units $\mu_0 = 4\pi \times 10^{-7}$ H/m (exactly), the $\varepsilon_0$ or $b$ must be determined by an independent experiment enabling a direct comparison of electric and magnetic forces acting at a distance or a comparison of units which are related to these forces.

Till now we have not used, with full awareness, the considerations belonging to the Special Theory of Relativity (STR). In spite of it we obtained conclusions and formula (7) which are, with the proviso that $b \equiv c$, traditionally attributed exclusively to the STR [6]. For example, it is stressed in many textbooks that the magnetism is a purely relativistic effect operating even at negligibly small relative velocities. Argumentation in favor of this statement is mostly based on the fact that it is possible, by means of kinematic principle of relativity, to derive from the electrostatic Coulomb law Ampère's law (or its equivalent the Biot-Savart law) and the Faraday induction law as well [7-9]. Nevertheless, as we have already seen above, the conclusion that the magnetism is a consequence of the relative motion of electric charges with respect to the observer must be more general, because it can be made without any reference to the postulates of the STR. Moreover, in the said relativistic derivations based on the Coulomb law its intrinsic element, the immediate *actio in distans*, was tacitly used, i.e. an assumption which is absolutely at odds with the gist of the STR, which is conceptually a field theory postulating the finiteness of the speed of interactions.

The effect of non-instantaneous interaction is, however, treated also in the frame of classical pre-maxwellian theories, e.g. in [10] or especially in excellent Riemann's posthumous paper [11] and in a more advanced and complete form in papers of Liénard and Wiechert [12,13], appearing already at the turn of the century. These authors assumed that the electric potential or electric forces acted with a delay corresponding to the transmission of the signal over the distance R. The actual time of interaction is thus given by a relation

$$t = \tau + R/b, \qquad (8)$$

where $\tau$ is the local time at the source of the electric force and $b$ is the speed of electromagnetic interaction. The potentials generated in this way are called retarded (Liénard-Wiechert's) potentials. Accordingly, instead of the distance $R$ the product $KR$ must be inserted into formulae (2) - (4), where $K$ is the correcting "Doppler factor" given by [14]

$$K = \partial t/\partial \tau = 1 + (1/b)\partial R/\partial \tau = 1 - (\mathbf{v} \, \mathbf{R_0})/b, \qquad (9)$$

$\mathbf{v}$ is the velocity vector and $\mathbf{R_0}$ the unit vector in the direction of the considered field point in the rest system. Let us apply this rule to the Coulomb law (2) using a somewhat special assumption that the vector $\mathbf{v}$ is parallel with the vector $\mathbf{R_0}$. The Coulomb law corrected by such a special Doppler factor then reads

$$F_{EM} = -(Q_1 Q_2 / 4\pi\varepsilon_0 R^2)(1 - v^2/b^2), \qquad (10)$$

where the symbol $F_{EM}$ is used for the retarded Coulomb force. Obviously, the physical content of this formula is essentially the same as that of formula (7). Interestingly enough, formula (10) is also identical with a central relationship of pre-maxwellian Wagner's electrodynamics (written for the particular case of zero acceleration) [10]. According to this theory $b$ represents a speed limit at which the electric force is just compensated by the invoked magnetic force (*"... die Constante b stellt dabei diejenige relative Geschwindigkeit vor, welche die elektrischen Massen $Q_1$ und $Q_2$ haben und behalten müssen, wenn gar nicht mehr auf einander wirken sollen."* [10]; as we are convinced, this old idea, i.e. that the electromagnetic field decouple from electric charges in cases where its relative speed attains value $b$, has an underestimated importance).

Regardless of the fact that we have derived the "relativistic" formulae (7) and (10) "classically" in a very elementary way and under rather special simplifying assumptions, it became clear that the existence of the magnetic interaction cannot be treated exclusively as a relativistic effect as is usually done. In our view, a more adequate statement is that the STR accounts for the magnetism just because it belongs to a wider class of theories implicitly involving the concept of non-instantaneous interaction. Notice in this connection that the experimentally established phenomenological laws (3) and (4), formulated in terms of instantaneous *actio in distans*, do not pretend to explain the magnetism but only to describe the real observations made in the rest-system. Therefore, although the algebraic operations with formulae (3) and (4) lead to relationship (7), they are not directly applicable to the systems moving relatively to the rest-system.

Let us now turn attention to the properties of the electromagnetic field decoupled from ponderable matter.

### 3. Maxwell's equations of the electromagnetic field

The structure of a hypothetic entity, electromagnetic field in an empty space, is controlled by reduced Maxwell's equations derived from experimentally observed laws mentioned above. In bi-vector SI notation they can be written in marvelously symmetric form due to Silberstein [15]:

$$\text{rot } \mathbf{\Lambda} = (i/b) \, \partial \mathbf{\Lambda}/\partial t, \tag{11}$$

$$\text{div } \mathbf{\Lambda} = 0, \tag{12}$$

where the electromagnetic complex bi-vector is defined as

$$\mathbf{\Lambda} = \mathbf{E} + ib\mathbf{B}, \quad i = \sqrt{(-1)}. \tag{13}$$

From these equations a lot of interesting conclusions can be made on the basis of purely mathematical deductions. For example, it can be proved that the discontinuity in the electromagnetic field is, in the absence of material bodies, either longitudinal stationary or constitutes a transversal vortex wave propagating with the velocity $b$ [16]. Since the theory based on equations (11) - (13) describes only the field itself (the so called *reine elektromagnetische Wellen,* i.e. "pure electromagnetic waves" by Silberstein) and not the interaction with ordinary matter, the origin of such discontinuities is out of the scope of the theory. It cannot be thus e.g. concluded that the electromagnetic wave excited by an oscillating electric dipole spreads from the source with velocity $b$. In many cases, namely, the speed of energy transfer represented by the flux of Poyinting's vector is appreciably slower than $b$

[17,18]. The transfer speed coincides with *b* only in the case of "purely electromagnetic" conjugate waves for which identically

$$\mathbf{E}^2 = b^2 \mathbf{B}^2. \tag{14}$$

However, condition (14) takes place practically only in the radiation zone, i.e. in the region sufficiently departed from all material bodies (sources, lenses, mirrors etc.). It can be stated quite generally that the speed of discontinuity of the electromagnetic field in the vicinity of ponderable matter is diminished and the speed *b* can be achieved only if the electromagnetic field is sufficiently decoupled from the material objects. As we believe, just this effect might be responsible for systematically lower observed values of the speed of light in cases where the microwaves or millimeter waves were used (see Fig. 1). The relative extent of zones where condition (14) is not exactly fulfilled is, namely, appreciably larger than that in experiments using visible light.

Nevertheless, just on the basis of the substantial agreement between the magnitudes of *c* and *b* Maxwell proposed (not without skepticism and in understandable ignorance of the facts just mentioned!) his famous "Dynamical Theory of the Electromagnetic Field" [19]. However, A. Einstein has gone much farther when, by postulating that in all inertial systems the kinematic "*speed of the ray of the light in vacuum is constant, being independent of movement of emitting body*" (Second postulate of STR, [20]), implicitly assumed that *c* is a universal constant identical with Maxwell's constant *b*, i.e.

$$c \equiv b. \tag{15}$$

This relation, considered in the frame of the STR as self-evident, is now regarded to be an independent postulate, called *c-equivalence principle* [21]. This principle has never been checked experimentally with sufficient accuracy, and therefore it bears somewhat philosophical character. Notice that the last electromagnetic measurements of Maxwell's constant *b* which were carried out in the year 1907 (asterisk in Fig. 1, [3]) provide a value of *b* differing essentially from that of *c*; (*b* = 299 710 000 ± 20 000 m/s, while exact value of *c* = 299 792 458 m/s). Latter, due to the general acceptance of the STR and of its Second postulate the experimentalists concentrate their efforts rather on the accuracy of kinematic speed of light measurements while the difficult and not very exact electric measurements of *b* were considered as pointless. It has to be stressed here, however, that such experiments had nothing to do with light and that *b* does not primarily represent the speed of light [22]. Restrained distinction between the electromagnetic waves and the light is felt also from Maxwell's comment concerning the nature of experiment for the determination of constants *b* and *c* [19]: "The value of *b* was determined by measuring the electromotive force with which the condenser of known capacity was charged, and then discharging the condenser through a galvanometer, so as to measure the quantity of electricity in it in electromagnetic measure. *The only use made of light in the experiment was to see the instruments.* The value of *c* found by M. Foucault was obtained by determining the angle through which a revolving mirror turned, while the light reflected from it went and returned along a measured course. *No use whatever was made of electricity and magnetism.*"

Maxwell's constant *b* being the only parameter controlling the behavior of pure electromagnetic field in vacuum is of primary significance in all considerations concerning this important physical entity. We are convinced that as such, it cannot be identified on the basis of more-or-less justified conjectures with another quantity (the speed of light, *c*-equivalence principle) without a sufficient experimental confirmation. However, modern measurements of the constant *b* having comparable accuracy as that of the speed of light are lacking. The knowledge of this constant would thus either fill the logical gap in the present theory or will provide a new experimental material for its further development.

### 4. Transformation properties of the electromagnetic field

Of special interest and profound physical significance are the transformation properties of reduced Maxwell's equations (11)–(13) describing pure electromagnetic field. As was convincingly shown by H. Bateman [23], reduced Maxwell's equations are invariant with respect to the group of spherical wave transformations which are in fact a generalized evocation of Huygens' principle. Unifying in a way usual in the STR temporal and spatial coordinates into Minkowski's space of four dimensions, it has further been shown that the spherical wave transformations can be there represented by a 15-element conformal mapping group. This group, actually identical with that studied by Sophus Lie, provides 15 so called Bessel-Hagen invariants [24]. It was shown only latter that the number of group elements and corresponding independent invariants could be reduced only to 11 [25]. However, even after such a reduction the number of elements of this group exceeds the number of elements in the group of the Lorentz transformations as introduced in the frame of the STR [14,26]. The group of these transformations known from the mathematics as "*la transformation par directions*

*réciproques*" differs from that of Sophus Lie (or Bessel-Hagen) in fact only by one element, namely, by the scale transformation. The very physical meaning of this transformation can be specified as independence of physical processes in the electromagnetic field on the absolute scale or dimensions in which these processes take place or, alternatively, as the absence of an intrinsic length in the electromagnetic field. In contrast to the pure electromagnetic field, in the world of ponderable matter the intrinsic length scales do exist; recall e.g. the Compton length of the electron which is $\lambda_C = 2\pi\hbar/m_e c$. It is probably a fundamental fact that the existence of intrinsic length scale in any system is conditioned by the presence of ponderable matter [27]. Recall in this connection Mach's conjecture claiming that "the matter creates the space", particularly that it defines shapes, points, etc., so that the operative geometry cannot do without material bodies. On the other hand, any "as if" entity, such as electromagnetic field, is for such a purpose quite insufficient. From this point of view the CGPM (1983) definition of meter sounds somewhat paradoxically; the unit of length is based on the entity for which the concept of length is absolutely irrelevant.

Comparing now the physical content of reduced Maxwell's equations of electromagnetic field with that established within the frame of the STR, one can immediately recognize that the limitations due to the introduction of the Second postulate of the STR, from which the group of Lorentz transformations takes its origin, are quite serious. The Second postulate eliminating the scale transformations thus consequently excludes the existence of some important Bessel-Hagen invariants, such as the analogue of spin of the Coulomb field. This prevents, e.g. to come to some results of Dirac's quantum theory independently. It should be also noticed that the existence of invariance of some Maxwell's equations under the time-independent conformal mappings was already discovered and used long time ago for the solution of electrostatic problems (cf. [5]). From this point of view the introduction of the Lorentz transformations into the theory must be regarded as a kind of compromise usable simultaneously for ponderable matter and electromagnetic radiation, mechanics and optics [26]. Unfortunately, such a compromise leading to a strict exclusion of some potential properties of the electromagnetic field from theoretical considerations must be classified as inconsistent.

## 5. Kinematic considerations

Prior to the discussion of kinematic aspects of the speed of light, it is necessary to distinguish between the so called two-way and one-way speed of light [28]. In the two-way speed time-of-flight experiment one must first trace out a closed path for the ray of light. By clocks placed at the starting point it is then measured the time during which the front of the light ray returns back. The ratio of the length of the closed path to the return time then defines the two-way speed of light. The time-of-flight measurement of the one-way speed of light, and, of course of one-way speed of any other physical entity, requires an accurate synchronization of the pair of clocks placed at the initial and terminal points of the testing path which is topologically a line segment. However, for the realization of the synchronization procedure either as prescribed by the STR [26,28-31], or by using "infinitesimally slow" transfer of clocks [33], it is necessary *a-priori* to assume the isotropy and the constancy of the one-way speed of light. Obviously, such a clear case of circular reasoning can only hardly be cured by a convention giving to the one-way speed of light a certain exact value, such as (1). It would be more reasonable to admit as an experimental fact that the measurement of one-way speeds is principally impossible and that the only two-way time-of-flight speed measurement can be realized using a closed testing path, which can do without synchronization of departed clocks. From the epistemological point of view thus the one-way speed of light must be excluded from our considerations and the Second postulate of the STR, in the above form, should be abandoned as an empty proposition. The difficult idea of the light spreading from the source into all directions with the same (one-way) constant speed $c$ regardless of inertial system from which it is observed as introduced by the Second postulate of the STR, will thus be replaced by a concept of a light which returns along the closed path (loop) to the starting point within the time just corresponding to the ratio = (length of the loop/$c$). Of course, in such a case one can easily imagine that the speed of light in moving inertial systems may differ in various points of space and directions, but only in such a way that the time-of-flight along all isometric loops would be the same because of some kind of "compensation" during the backward movement. In other words, in such a version of the Second postulate it is assumed that *the two-way speed of light measured along different loops in different inertial frames would represent exactly the same constant.* Simultaneously, it is just the condition restricting a set of admissible space-time transformations reformulated in terms of two-way speed of light. Making such an assumption, it is among others immediately clear that the Michelson – Morley experiment in which a split light ray is conveyed along two isometric loops has to provide a well known "negative" result, or, directly confirms the "two-way speed version" of the Second postulate.

Basing on these ideas, it starts to be apparent that the most problematic operation in measuring speeds is the synchronization of clocks at departed stations. Therefore, let us mention here the standard technique for the determination of one-way delay time $t_A$ necessary for propagation of light signal from point "1" to another point "2" which seemingly circumvents the synchronization of the clocks [32]. (This technique was also used e.g. in famous OPERA experiment [34] where superluminous speed of neutrino was reputedly indicated.) Split light signal is sent from point "1" via two alternative paths A (direct) and B (slightly longer) to point "2", where the time difference $t_A - t_B$ is measured between arrivals of both signals. After that the total time $t_A + t_B$ is measured which is necessary for the propagation of the signal from "1" to "2" along the path A and back to point "1" along the path B. The time $t_A$ which we are interested in, can be then computed by an obvious formula

$$t_A = \tfrac{1}{2}\,[(t_A + t_B) + (t_A - t_B)]. \qquad (16)$$

Evidently the method is fully based on the assumption that the time $t_B$ needed for the signal propagation along the path B from "1" to "2" is exactly the same as that for backward propagation from "2" to "1". Such an assumption is, however, equivalent to the validity of the Second postulate of STR, a statement for which the experimental evidence is in fact lacking. Admitting instead of the two-way speed version of the Second postulate confirmed by experiments, the times corresponding to the signal propagation in opposite directions may generally differ, i.e. $t_B(1\rightarrow 2) \neq t_B(2\rightarrow 1)$, which makes formula (16) useless. In such a case the results of speed measurements based on it will be inevitably spurious. This example clearly illustrates that the Second postulate of STR is not only redundant but that it can be also harmful.

    The redundancy and arbitrariness of the Second postulate of the STR was recognized by a group of Chinese researchers who formulated a theory of relativity which can do without Second postulate intimately related to the speed of light [35]. This is so called "Taiji relativity". (The term "Taiji" is a central concept of ancient Chinese cosmology having a meaning of ultimate principle existing before the creation of the Universe. As such, it is akin to the European term apeiron - άπειρον.) It claims among others that the relativity of time and the universality of speed of light are not physical entities inherent in nature but human conventions imposed upon it. The Taiji theory is in fact essentially based on the experimentally confirmed two-way speed version of the Second postulate of the STR as formulated above, according to which the two-way speed must be isotropic in all inertial frames.

    The fact that only the two-way speed of light has physical meaning has its consequences also for the realization of the meter according to its CGPM (1983) definition.

The authors of this definition and recommendations for the practical realization of the meter normal (*mise en pratique* [2]), namely, tacitly count with the possibility of one-way measurement of the speed of light and with their full awareness admit the postulates of the STR. The shape of the test path need not be thus specified and the time measurement, particularly the synchronization of clocks, is done according to the "accepted good practice" which follows the rules of STR. Astonishingly, it is further assumed that the measurements are restricted to the lengths which are sufficiently short to be negligible for the effects belonging to the scope of general relativity (e.g. gravitation). Such an assumption logically challenges the fact that the universal constant *c* is "universal" and "constant". It is thus apparent that the critical reconsideration of the definition and of allied problems is necessary.

Taking into account the fact that the most accurate arrangements for the optical measurements of the speed of light operate as a two-way laser interferometers of Fabry-Perot type where the test path is enclosed between two parallel semi-permeable mirrors, the normal of length can be quite naturally realized just by the distance between these two plains. However, the distance defined in this way can be converted into terms inherent to the CGPM (1983) definition only by a consequent use of the Second postulate of STR. The CGPM (1983) definition thus makes this postulate conceptually indispensable although it has for the practical realization of the normal of length no significance.

## 6. G. de Bray scenario

As was already mentioned above, according to the CGPM (1983) definition of meter and to the corresponding *mise en pratique*, the measurements should be restricted to the lengths which are "sufficiently" short in order not to be influenced by gravitation etc. Beside the space-related effects, however, there can possibly also be the time drift of the value of *c*. To this effect turned attention G. de Bray who after the careful revision of experiments made by various researchers between the years 1849-1933 concluded that *the earlier observations give a higher speed of light* [36]. Such a behavior of the "pseudoconstant" *c* has been attributed by the author to the apparent decrease of the unit of length caused by the expansion of the Universe. If the radius of the Universe doubles every $K$ years, then the measured velocity of light will be halved every $K$ years, that is, the velocity will be proportional to $c \propto (1/2)^{\tau/K}$, where $\tau$ is here time in years. Of course, an enormous increase of accuracy of measurements and the constancy of the speed of light during the second half of the 20[th] century made the G. de Bray scenario highly improbable. Nevertheless, to be correct, the long-term drift of *c* can be hardly excluded or proved on the basis of say ~10 measurements made during two decades only (see Fig. 1). Since the influence of cosmological effects belonging to the scope of general relativity cannot be *a-priori* excluded even for terrestrial observations, the CGPM (1983) definition anticipating in fact the absolute impossibility of G. de Bray scenario sounds rather bold and not very wise. Moreover, as we have seen above, the CGPM (1983) definition is not a plain technicality but it contains hidden assumptions which can make the analysis of experiments confused and difficult.

## 7. Conclusions

In the present contribution we have discussed, as we believe, a legitimate question, what the consequences of the CGPM (1983) definition of meter are, especially of the fixation of the speed of light at the exact value $c = 299792458$ m/s. This question has important aspects which, remaining in the shade of the STR, are discussed in the current literature only marginally. However, as we are convinced, their omission may lead and led to the incorrect conclusions or to the masking of real effects. Therefore, these aspects should be analyzed critically and repeatedly. Some of them were discussed in the present paper with the following conclusions.

First of all, we have summarized the arguments in favor of the idea that the magnetic force is a manifestation of the electric force in relative movement. We have shown that such a picture is independent of STR and of its Second postulate and that the square of Maxwell's constant $b$ controlling the ratio of electric and magnetic force in any particular situation is not necessarily related to the speed of light, $c$.

Although we are not convinced that the $c$-equivalence principle is not valid ($c \neq b$), we have stressed that its experimental confirmation is at present lacking. Consequently, there is a serious logical gap in the electromagnetic theory of light and all related theories as well. According to our meaning, it is thus quite necessary to repeat the measurements of $b$ with the accuracy comparable with that of the speed of light in the near future

Considering the description of the electromagnetic field, we have turned our attention to the fact that the transformation group of Maxwell's equations is larger than the Lorentz group, in other words that there are invariants (Bessel-Hagen invariants) and properties of the electromagnetic field that are out of the scope of the STR. This fact may have a lot of important consequences which should not be omitted only on the basis of a teleological argument, i.e. because they do not fit well the STR.

In the paragraph concerning the kinematic problems we have pointed out the principal impossibility of measurement of the one-way speed of light and other physical entities. Therefore, we have suggested a "two-way speed" version of the Second postulate of the STR which fits well the known experimental observations and in some cases effectively eliminates the difficulties with synchronization of clocks. Some features of this approach are common with the so called Taiji theory of relativity which left out the Second postulate of the STR completely. As we have further shown the ignorance of the principal impossibility of measurement of the one-way speeds may lead to serious experimental errors and misinterpretations.

The CGPM (1983) definition anticipates the absolute impossibility of the scenario of G. de Bray's type, i.e. of slow drift of the speed of light, although such a drift is compatible with the predictions of e.g. general theory of relativity. We are therefore convinced that the measurements of the speed of light should remain the subject of astrophysical experimental research in future, in spite of the fact that they already lost their formal sense.

And at last but not least, although we criticized, in this contribution among others, one of the corner-stones of modern physics, the Second postulate of the STR, our aim was by no means to belittle the merits and great efforts of scientists such as W. Voigt, J. Larmor, H. A. Lorentz, H. Poincaré and A. Einstein who established the STR more than one hundred years ago but only to turn attention to the problems which should be solved. Let us therefore add a quotation from the paper of one of the creators of the STR (H. Poincaré [32]) which elucidates also our attitude: *"Good theories are flexible. ... Specious arguments have no effect on them, and they also triumph over all serious objections. However, in triumphing they may be transformed."*